\def\cao{\c c\~ao\ }

\def\1{\'{\i}}
\def\beq{\begin{equation}}
\def\eeq{\end{equation}}
\def\bea{\begin{eqnarray}}
\def\eea{\end{eqnarray}}
\def\bed{\begin{displaymath}}
\def\eed{\end{displaymath}}
\def\no{\noindent}
\def\nn{\nonumber}

\documentclass[preprint,pre]{revtex4-1}
\usepackage{graphics,graphicx,amssymb,amsmath,dcolumn,bm}

\begin{document}

\title{Turing instability in oscillator chains with non-local coupling}

\author{R. L. Viana \footnote{Corresponding author. e-mail: viana@fisica.ufpr.br}, F. A. dos S. Silva, and S. R. Lopes}
\affiliation{Departamento de F\1sica, Universidade Federal do Paran\'a, Caixa Postal 19044, 81531-990, Curitiba, Paran\'a, Brazil.}

\begin{abstract}
We investigate analytically and numerically the conditions for the Turing instability to occur in a one-dimensional chain of nonlinear oscillators coupled non-locally in such a way that the coupling strength decreases with the spatial distance as a power-law. A range parameter makes possible to cover the two limiting cases of local (nearest-neighbor) and a global (all-to-all) couplings. We consider an example from a non-linear auto-catalytic reaction-diffusion model.
\end{abstract}
\date{\today}

\maketitle

\section{Introduction}

Given the diversity of spatial patterns one observe in Nature, it is always puzzling to wonder how this diversity arises from spatially uniform situations. This question was pursued by A. Turing in his seminal 1950's paper on the mechanism of morphogenesis \cite{turing}, in which he considered how can a spatially homogeneous state lose stability, giving origin to a spatially non-homogeneous pattern, the so-called Turing instability. Turing analyzed a reaction-diffusion system comprised of spatially coupled ordinary differential equations that modelled discrete biological cells. 

Turing put forward two basic assumptions: (i) the need of at least two chemical substances (morphogenes), an activator and an inhibitor; (ii) diffusion plays a destabilizing role in the interacting chemical substances. The latter point is somewhat counterintuitive, for diffusion usually smooths out spatial structures in linear systems and thus would be expected to exert a stabilizing influence instead. Moreover, Turing found that the instability caused by diffusion leads to the growth of spatial structure at a particular wave length. This mechanism has explained successfully a wealth of pattern formation phenomena in Nature, like the segmentation patterns in the developing fly embryo, the periodic array of tentacles around the mouth of the Hydra, zebra stripes \cite{murray}, sea shell striations \cite{meinhardt4}, to name just a few. In physical chemistry Turing instability has been found in many continuously fed stirred tank reactors, like the Chlorine Dioxide-Iodine-Malonic Acid reaction \cite{lengyel,cross}.

From the mathematical point of view, Turing instability arises when an otherwise stable homogeneous state becomes unstable as a result of diffusion, in the sense that any small perturbation will develop a spatially nonuniform state. Linear theory can predict the wave length of the unstable mode which grows \cite{atlee}. This growth, however, is limited by the saturating effect of nonlinear terms and a spatially non-homogeneous pattern is then formed as a result of Turing instability. 

Turing instability is usually investigated in the context of local couplings, in which each cell interacts with its nearest-neighbors only. This is modelled by Fick's law, in which the diffusive flux of morphogenes (activator and inhibitor) is proportional to their local concentration, being responsible for the second spatial derivatives in the mathematical models of such reaction-diffusive systems. However, there are many situations in which this picture is over-simplified, like in chemical couplings between biological cells, generating non-local interactions which cannot be described by a diffusive coupling. Non-local couplings can be implemented in macroscopic density equations in which the diffusion coefficient depends on a weighted spatial average of the density \cite{lopez}. 

Such nonlocal couplings take into account not only the nearest neighbors of a given cell, but also the other cells of the assembly. An extreme form of this arises when each cell couples with the mean field produced by all the other cells, the so-called global (all-to-all) coupling, which is used in models of neural networks \cite{pikowsky}, coupled Josephson junctions \cite{strogatz}, and it is the basis of the paradigmatic Kuramoto model \cite{kuramoto}. It is possible to interpolate between nearest-neighbor and all-to-all coupling by considering an interaction term whose strength decreases with the physical distance between cells as a power-law \cite{rogers}. This model has been used to theoretical and numerical studies of various spatio-temporal phenomena like synchronization, shadowing and cluster formation \cite{nosso}.

The influence of non-local couplings on the formation of Turing patterns has been studied in many systems. The conditions for the occurrence of Turing instability were investigated for logistic growth processes using various non-local coupling terms \cite{shnerb}. In other recent study, a three-variable Oregonator model of a light-sensitive Belousov-Zhabotinsky reaction was studied, in which the nonlocal coupling was externally imposed by an optical feedback loop \cite{nicola}. In the latter investigation, it was found that long-range inhibition leads to Turing instability, whereas long-range activation induces wave-like patterns. 

This work aims to investigate the presence of Turing instabilities in oscillator chains with a non-local coupling of the power-law form, so as to have results which hold both for local and global cases. We derive analytically the conditions for a Turing instability to occur in a one-dimensional lattice of two-dimensional systems (representing the activator-inhibitor pair) undergoing linear dynamics. Then we study a nonlinear reaction-diffusion system proposed by Meinhardt and Gierer \cite{meinhardt1} as a model for pattern formation related to skin pigmentation. 

The rest of this paper is organized as follows. In Section II we introduce the oscillator model to be studied, as well as the type of coupling between the units. The analytical conditions for a Turing instability to occur are derived in Section III using linear stability theory. The effect of nonlinear terms and the ensuing pattern formation is discussed in Section IV through the Meinhardt-Gierer model. Our conclusions are left to the last Section.

\section{Oscillator chain with non-local coupling}

Let us consider an oscillator one-dimensional chain with $N$ units, each of them being a nonlinear reaction-diffusion system, whose dynamical state is described by the concentrations of the activator $x_k(t)$ and inhibitor $y_k(t)$, satisfying the following equations ($k = 1, 2, \cdots N$):
\bea
\label{chainx}
{\dot x_k} & = & {\bf X}(x_k,y_k) - D_x x_k + \frac{D_x}{\kappa(\alpha)} \sum_{r=1}^{N'} \frac{1}{r^\alpha} \left( x_{k-r} + x_{k+r} \right), \\
\label{chainy}
{\dot y_k}  & =  & {\bf Y}(x_k,y_k) - D_y y_k + \frac{D_y}{\kappa(\alpha)} \sum_{r=1}^{N'} \frac{1}{r^\alpha} \left( y_{k-r} + y_{k+r} \right), 
\eea
\no where ${\bf X}$ and ${\bf Y}$ stand for the vector field corresponding to the uncoupled oscillators. We use a non-local form of coupling for which the interaction strength decreases with the lattice distance in a power-law fashion, where $D_x$ and $D_y$ are positive coupling constants, representing the different diffusion coefficients of the chemical species, $\alpha$ is a positive real number, and
\beq
\label{kappa}
\kappa (\alpha) = 2 \sum_{r=1}^{N'} \frac{1}{r^\alpha},
\eeq
\no is a normalization factor, with $N' = (N-1)/2$, supposing that $N$ is an odd number. In the oscillator chain given by Eqs. (\ref{chainx})-(\ref{chainy}), the summation terms are weighted averages of discretized second spatial derivatives, the common normalization factor $\kappa(\alpha)$ being the sum of the corresponding weights. 

In the limit $\alpha \rightarrow \infty$, only the $r = 1$ term will survive in the summation terms, resulting in $\kappa \rightarrow 2$ and the Laplacian or diffusive coupling 
\bea
\label{chainxd}
{\dot x_k}  & =  & {\bf X}(x_k,y_k) + \frac{D_x}{2} \left( x_{k-1} - 2x_k + x_{k+1} \right), \\
\label{chainyd}
{\dot y_k}  & =  & {\bf Y}(x_k,y_k) + \frac{D_y}{2} \left( y_{k-1} - 2y_k + y_{k+1} \right),
\eea
\noindent in which only the nearest-neighbors of a given site contribute to the coupling term. The other limiting case, $\alpha = 0$, is such that $\kappa(0) = N - 1$ and the oscillator chain becomes globally coupled 
\bea
\label{chainxg}
{\dot x_k}  & =  & {\bf X}(x_k,y_k) + D_x \left( - x_k + \frac{1}{N-1} \sum_{r=1,r\ne k}^{N} x_r \right), \\
\label{chainyg}
{\dot y_k}  & =  & {\bf Y}(x_k,y_k) + D_y \left( - y_k + \frac{1}{N-1} \sum_{r=1,r\ne k}^{N} y_r \right),
\eea
\noindent where each oscillator interacts with the mean value of all lattice sites, irrespective of their positions (``mean-field'' model). Hence, the coupling term in Eqs. (\ref{chainx})-(\ref{chainy}) may be regarded as an interpolating form between these limiting cases, and will be referred to as a {\it power-law coupling}.

\section{Linear stability}

Let us consider the equilibrium values of the activator and inhibitor given, respectively, by $X_0$ and $Y_0$. The deviations from these values (also denoted by $x$ and $y$, for notational simplicity), in lowest order, obey a linearized model, for which the activator-inhibitor dynamics in each uncoupled oscillator (labelled by $k$, as before) is governed by the affine vector field 
\bea
\label{activator}
{\bf X}(x_k,y_k) & = & a x_k + b y_k, \\
\label{inhibitor}
{\bf Y}(x_k,y_k) & = & c x_k + d y_k, 
\eea
\no where the constants $a$ to $d$ are the elements of the Jacobian matrix of the nonlinear vector field, evaluated at $(X_0,Y_0)$. The equilibrium solution of the linearized system is $(x^*=0,y^*=0)$, which is linearly stable if 
\beq
\label{equiun}
q \equiv ad-bc > 0, \qquad {\mbox{\rm and}} \qquad a+d < 0.
\eeq
\no The convergence to the equilibrium will be monotonic if ${(a-d)}^2 +4bc > 0$ and oscillatory if this factor is negative. This set of conditions shall be supposed from now on.

\begin{figure}
\includegraphics[width=1.0\columnwidth]{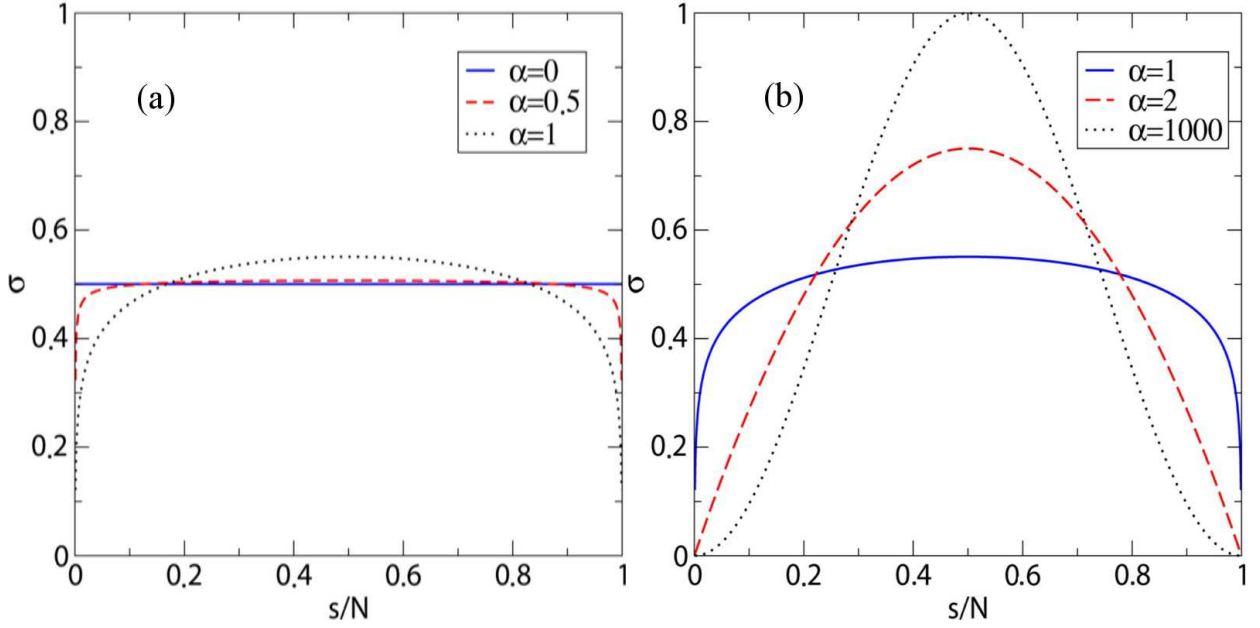}
\caption{\label{sigmafig} Dependence of the function defined by Eq. (\ref{sigma}) on the argument $s/N$, for (a) small and (b) large values of the parameter $\alpha$.}
\end{figure}

In order to treat the coupled system we can use a discrete Fourier transform of each dynamical variable, in the following form
\bea
\label{fourierx}
x_k(t) & = & \sum_{s=0}^{N-1} \xi_s(s,t) \exp\left(\frac{2\pi i s k}{N}\right), \\
\label{fouriery}
y_k(t) & = & \sum_{s=0}^{N-1} \eta_s(s,t) \exp\left(\frac{2\pi i s k}{N}\right),
\eea
\no where $\xi_s$ and $\eta_s$ are time-dependent Fourier mode amplitudes. On substituting (\ref{fourierx})-(\ref{fouriery}) in the linearized version of (\ref{chainx})-(\ref{chainy}) we find that the mode amplitudes satisfy uncoupled linear differential equations:
\bea
\label{chis}
{\dot\xi_s} & = & [a - 2 D_x \sigma_\alpha(s,N)] \xi_s + b \eta_s = a_\sigma \xi_s + b \eta_s , \\
\label{etas}
{\dot\eta_s} & = & [d - 2 D_y \sigma_\alpha(s,N)] \xi_s + c \eta_s = c \xi_s + d_\sigma \eta_s , 
\eea
\no where we define the following auxiliary quantities:
\bea
\label{asigma}
a_\sigma & \equiv & a - 2 D_x \sigma_\alpha(s,N), \\
\label{dsigma}
d_\sigma & \equiv & d - 2 D_y \sigma_\alpha(s,N), \\
\label{sigma}
\sigma_\alpha(s,N) & \equiv & \frac{1}{2} - \frac{1}{\kappa(\alpha)} \sum_{r=1}^{N'} \frac{1}{r^\alpha} \cos\left(\frac{2\pi sr}{N}\right)
\eea
\no is a function whose values are such that $0 \le \sigma_\alpha(s,N) \le \sigma_{max}(\alpha,N) \le 1$ [Fig. \ref{sigmafig}]. For the nearest-neighbor case ($\alpha\rightarrow\infty$) it turns out that 
\beq
\label{sigmainf}
\sigma_\infty(s,N) = \sin^2 \left(\frac{\pi s}{N} \right),
\eeq
\no for which $\sigma_{max} = 1$ reaches its largest value [Fig. \ref{sigmafig}(b)]. On the other hand, in the mean-field coupling ($\alpha = 0$) this function reads
\beq
\label{sigma0}
\sigma_0(s,N) = \frac{1}{2} \left(\frac{N}{N-1}\right) (1 - \delta_{s,0}),
\eeq
\no such that, for large $N$ it tends to $1/2$ [Fig. \ref{sigmafig}(a)]. 

The Turing instability occurs when a spatially homogeneous pattern becomes inhomogeneous by virtue of the coupling effect which acts as a perturbation on each oscillator, driving it out of its equilibrium state. The equilibrium point of the linearized mode amplitude equations, $(\xi_s,\eta_s)=(0,0)$, corresponding to equilibrium concentrations of activator and inhibitor species for each oscillator. Hence the onset of a Turing instability happens whenever this equilibrium becomes an unstable saddle point, which occurs if
\beq
\label{qdet}
q_\sigma \equiv a_\sigma d_\sigma - bc = 4 D_x D_y  \sigma^2 - 2\sigma(a D_y - d D_x) + q < 0.
\eeq
\no On defining the auxiliary variables
\beq
\label{auxvar}
P \equiv \frac{a}{D_x} + \frac{d}{D_y}, \qquad
Q \equiv \frac{ad-bc}{D_x D_y},
\eeq
\no the inequality (\ref{qdet}) is satisfied provided $\sigma_- < \sigma < \sigma_+$, where
\beq
\label{sigmapm}
\sigma_{\pm} \equiv \frac{P \pm \sqrt{P^2-4Q}}{4}.
\eeq
\no Since the function $\sigma$ has the upper bound $\sigma_{max}(\alpha,N) \le 1$ we must have $0 \le \sigma_- \le \sigma_{max}(\alpha,N) $, such that $0 \le P - \sqrt{P^2-4Q} \le 4 \sigma_{max}(\alpha,N) $, which gives the following conditions for the appearance of Turing instability in the system
\bea
\label{cond1}
Q & > & 0, \\
\label{cond2}
P & > & 2 \sqrt{Q}, \qquad {\mbox{\rm if}} \qquad 0 \le P \le 4 \sigma_{max}(\alpha,N) , \\
\label{cond3}
P & > & \frac{Q}{2 \sigma_{max}(\alpha,N) } + 2 \sigma_{max}(\alpha,N) , \qquad {\mbox{\rm if}} \qquad P > 4 \sigma_{max}(\alpha,N) . 
\eea

\begin{figure}
\includegraphics[width=1.0\columnwidth]{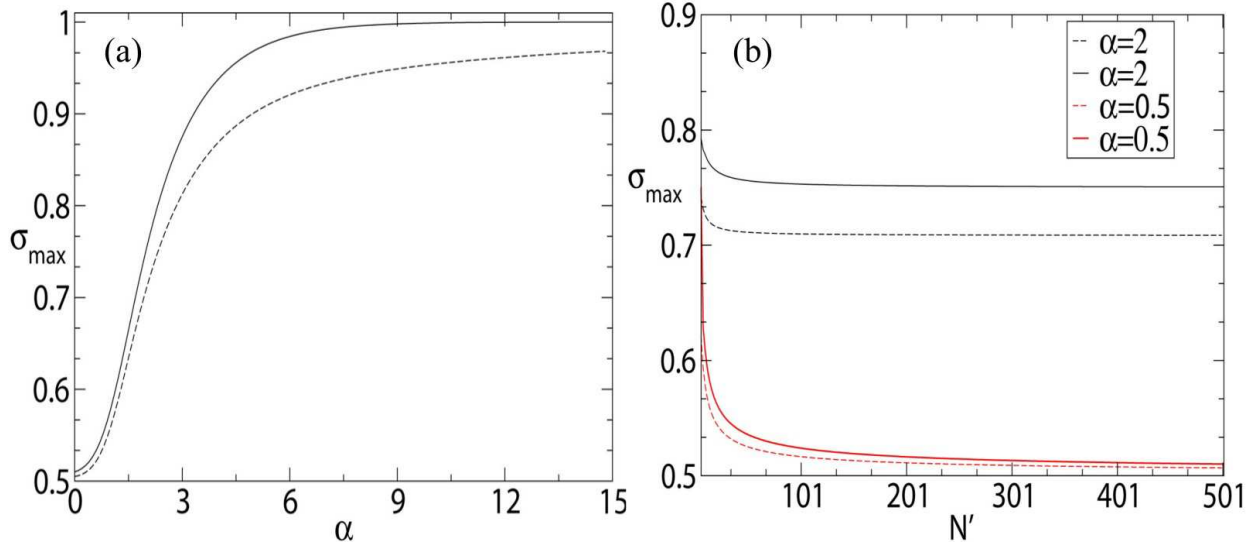}
\caption{\label{aproxfig} Dependence of the maximum value of $\sigma_\alpha(s,N)$ on the parameter $\alpha$. The full curve represents a numerical evaluation of the summation, whereas the dashed curve results from an analytical approximation.}
\end{figure}

Hence the stability of the spatially homogeneous state in this system depends on the values taken on by $\sigma_{max}$. We show in Fig. \ref{aproxfig} how it depends on the range parameter $\alpha$ by direct numerical evaluation of the function extremum (solid curve) and using an analytical approximation (dashed curve). The latter was obtained as follows: first we can show from (\ref{sigma}) that this function is always symmetric with respect to $s = N/2$:
\[
\sigma_\alpha(s=N/2+L,N) = \sigma_\alpha(s=N/2-L,N),
\] 
\no for all $L$ values such that $0 \le s \le N-1$. 

From the same token we show that $\sigma_\alpha(s,N)$ has a local maximum at $s = N/2$ (for $N'=(N-1)/2$ odd)
\[
{\left.\frac{\partial \sigma_\alpha(s,N)}{\partial s}\right\vert}_{s=N/2} = 0,
\qquad 
{\left.\frac{\partial^2 \sigma_\alpha(s,N)}{\partial s^2}\right\vert}_{s=N/2} < 0.
\]
\no such that
\[
\sigma_{max}(\alpha,N)  = \sigma_\alpha(s=N/2,N).
\]

Finally, if $N$ is large enough that $s/N$ can be reasonably well approximated by a continuous variable, we can use the Euler-MacLaurin formula \cite{abramowitz} to replace the summation in (\ref{sigma}) by an integral, which yields an analytical approximation for $\sigma_{max}$:
\bea
\nn
\sigma_{max}(\alpha,N)  & \approx & \frac{1}{2} - \frac{1}{2}{\left\lbrack
\frac{1-{(N')}^{1-\alpha}}{\alpha -1} + \frac{{N'}^{-\alpha}+1}{2} \right\rbrack}^{-1} 
\left\{
\left\lbrack- \frac{1}{2}
\left( \frac{1-{(N')}^{1-\alpha}}{\alpha -1} \right) - \left(\frac{{N'}^{-\alpha}+1}{2} \right) \right\rbrack + \right. \\
\label{analsigma}
& & \left.
\left\lbrack \frac{2^{-\alpha}}{\alpha-1} (1-2^{\alpha-1}{(N'-1)}^{1-\alpha}) + \frac{{(N'-1)}^{-\alpha} + 2^{-\alpha}}{2}
\right\rbrack
\right\},
\eea
\no which holds for odd values of $N'$. We see that the approximation always underestimate the actual value of $\sigma_{max}(\alpha,N) $, but the difference turns not to be large, specially at small $\alpha$. On the other hand, for large $\alpha$, while the analytical approximation gives worse results, we already know that this value must approach the unity, since this is the nearest-neighbor coupling case. Notice that, in this work, the large-$N$ limit has to be taken only as an approximation for obtention of closed-form analytical expressions. In fact, for $N\rightarrow\infty$ the power-law coupling (\ref{chainx})-(\ref{chainy}) is not normalizable when $0 < \alpha < 1$ \cite{ital}.

In the case of local (nearest-neighbor) coupling ($\alpha \rightarrow \infty$) we have large values of $\alpha$ and $\sigma_{max}$ tends to the unity, such that the conditions (\ref{cond1})-(\ref{cond3}) a Turing instability to occur reduce to 
\bea
\label{cond1loc}
Q & > & 0, \\
\label{cond2loc}
P & > & 2 \sqrt{Q}, \qquad {\mbox{\rm if}} \qquad 0 \le P \le 4, \\
\label{cond3loc}
P & > & \frac{Q}{2} + 2, \qquad {\mbox{\rm if}} \qquad P > 4,
\eea
\no which agrees with the results of Ref. \cite{atlee} up to a (nonessential) factor of $2$ due to a slightly different definition of the local coupling prescription. 

Since he uncoupled oscillators are supposed to have a stable equilibrium concentration, the local conditions (\ref{equiun}) apply, such that (\ref{cond1loc}) always holds. The remaining conditions depend on the values of the diffusion coefficients. It is instructive to consider the case where both diffusion coefficients are equal: $D_x = D_y = D$. In this case it turns out that $P = (a+d)/D$ is negative, and there cannot occur a Turing instability. In fact, it is a well-known fact that pattern formation in a chemical system will not occur unless the diffusion coefficients of the activator and inhibitor differ substantially \cite{cross}.

The range of unstable modes, in the local coupling case, is given by
\beq
\label{range0}
\min(\sigma_+,1) > \sigma > \sigma_-,
\eeq
\no where $\sigma_{\pm}$ are given by (\ref{sigmapm}). Notice that we have $\sigma_+ > 1$ provided $P < (Q/2) + 2$. Substituting (\ref{sigmainf}) and solving the resulting inequalities there result two intervals of unstable modes symmetrically located with respect to $s/N=1/2$:
\bea
\nn
\frac{\sin^{-1} \sqrt{\sigma_-}}{\pi} < \frac{s}{N} < \frac{\sin^{-1} \sqrt{\sigma_+}}{\pi} , \\
\label{range}
1 - \frac{\sin^{-1} \sqrt{\sigma_+}}{\pi} < \frac{s}{N} < 1 - \frac{\sin^{-1} \sqrt{\sigma_-}}{\pi} , 
\eea
\no In this way, a larger number of modes become unstable if $\sigma_+ > 1$.

The global (all-to-all) case ($\alpha=0$) presents interesting features, for the function $\sigma_0(s,N)$ takes on a constant value $\sigma_0$ given by (\ref{sigma0}). The conditions (\ref{cond1})-(\ref{cond3}) hence reduce to a single inequality
\beq
\label{condglo}
P > \frac{Q}{2\sigma_0} + 2 \sigma_0,
\eeq
\no in such a way that, once this condition is fulfilled, all modes become simultaneously unstable. Since, for large $N$, we have $\sigma_0 \approx 1/2$, the relation above can be written in a simple form $P > Q+1$. It is interesting to note that, if the diffusion coefficients are equal, this condition cannot be satisfied in general, as in the local case. Thus there is a range-independent influence of the dissimilarity of the diffusion coefficients on the occurrence of a Turing instability.

\begin{figure}
\includegraphics[width=1.0\textwidth]{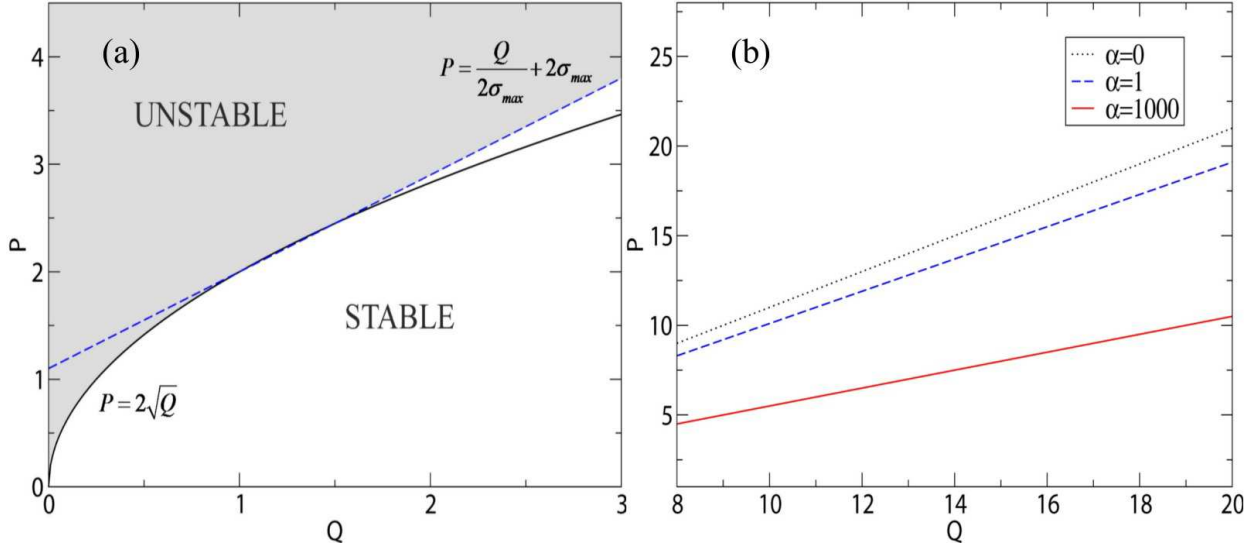}
\caption{\label{param} (a) Parameter plane showing the lines of marginal stability for the case of $N = 101$ coupled oscillators and arbitrary $\alpha$. (b) Magnification of a region of (a) of large $Q$-values}
\end{figure}

Let us consider a specific example of the intermediate range case, for which $\alpha = 1$ and we take a chain with $N = 101$ oscillators. Then (\ref{sigma}) gives $\sigma_{max} \approx 0.577$ for this case. The approximation expressed by Eq. (\ref{analsigma}), which works well also for $N'$ even, would give a slightly smaller value of $0.568$ for $\sigma_{max}$. On substituting the value of $\sigma_{max}$ in Eqs. (\ref{cond1})-(\ref{cond3}) we obtain conditions for the existence of a Turing instability which can be better represented in the parameter plane depicted (for $\alpha$ arbitrary) in Fig. \ref{param}(a), where we draw the lines where we have marginal stability as a function of parameters $P$ and $Q$. Since, for $\alpha = 1$, the function $\sigma$ has a broad flat top, failing on being a constant (like in the global case) only for the vicinity of the extremities $s/N=0$ and $s/N = 1$, we have roughly the same behavior in that most modes become unstable simultaneously in the intermediate range case. 

A useful information is the stable area $A_{st}$, or the area in the parameter plane for which we have stable equilibria. In Fig. \ref{param}(b) we show a magnification of a region of large $Q$-values in the parameter plane for different values of $\alpha$. It is apparent that the stable area decreases with increasing $\alpha$, i.e. the relative number of parameter values yielding stable equilibria is larger as the coupling becomes global, and smaller as the coupling becomes local. By integrating out the stability curves in Fig. \ref{param}(a), from $Q = 0$ to an arbitrary $Q_0$, it is actually possible to derive an analytical expression for the stable area (normalized by $Q_0^2$), which reads
\beq
\label{starea}
A_{st} = \frac{1}{4\sigma_{max}} + \frac{\sigma_{max} (2 Q_0 - 4)}{Q_0^2} + \frac{32\sigma_{max}^{3/2}}{3 Q_0^2} - 8 \frac{\sigma_{max}^2}{Q_0^2},
\eeq
\no giving, for large $Q_0$, $A_{st} \approx 1/(4\sigma_{max})$. In very long oscillator chains, where $N$ is large, with local and global couplings $\sigma_{max}$ takes on values equal to $1$ and $1/2$, respectively. Hence the stable area in the local case is roughly half of the corresponding area in the global case. We thus conclude that Turing instability would be statistically more common for local couplings than for global ones. 

\section{Nonlinear reaction-diffusion model}

In reaction-diffusion systems of practical interest, there are nonlinear terms which exert saturation on the non-homogeneous spatial modes excited by the Turing instability, giving rise to pattern formation. The basic mechanism of this saturation combines a short-ranged auto-catalytic activation, which causes self-enhancement after Turing instability sets in, with a long-ranged inhibitory effect. The latter can be caused by either a rapidly spreading and long-ranging or a result from a depletion of material required for the self-enhancement that is obtained from the surrounding region \cite{meinhardt1}. 

One paradigmatic example of this broad class of reaction-diffusion systems is the Meinhardt-Gierer model which, in the language of Eqs. (\ref{chainx})-(\ref{chainy}), reads \cite{meinhardt1,meinhardt2}
\bea
\label{mg1}
{\bf X}(x_k,y_k) & = & \rho_x \frac{x_k^2}{y_k} - \mu_x x_k, \\
\label{mg2}
{\bf Y}(x_k,y_k) & = & \rho_y x_k^2 - \mu_y y_k,
\eea
\no where $\rho_{a,h}$ and $\mu_{a,h}$ are positive parameters characterizing the local reaction rates. 

The term proportional to $\rho_x$ comes from the assumption that the activator is an auto-catalytic chemical species, its denominator representing the inhibitor effect of the other species. In this case $\rho_x$ turns to be a small activator-independent production rate of the activator and is actually necessary to start the activator auto-catalysis at very low activator concentration, like in the case of regeneration. Likewise, $\rho_y$ stands for a baseline production rate of the inhibitor \cite{meinhardt1}. The quantities $\mu_x$ and $\mu_y$ play the role of degradation of the activator and inhibitor species, respectively, such that the number of molecules of each species that decay per time unit is proportional to the corresponding decay rate and to the number of molecules themselves \cite{meinhardt3}. 

\subsection{Stability analysis}

We first determine the equilibrium points of the uncoupled vector field (\ref{mg1})-(\ref{mg2}). One of them is obviously $x^* = y^* = 0$, whereas there is a non-trivial equilibrium with activator and inhibitor concentrations given, respectively, by
\beq
\label{equil}
x^* = \frac{\rho_x \mu_y}{\rho_y \mu_x}, \qquad 
y^* = \frac{\rho_x^2 \mu_y}{\rho_y \mu_x^2}, 
\eeq

\begin{figure}
\includegraphics[width=0.7\columnwidth]{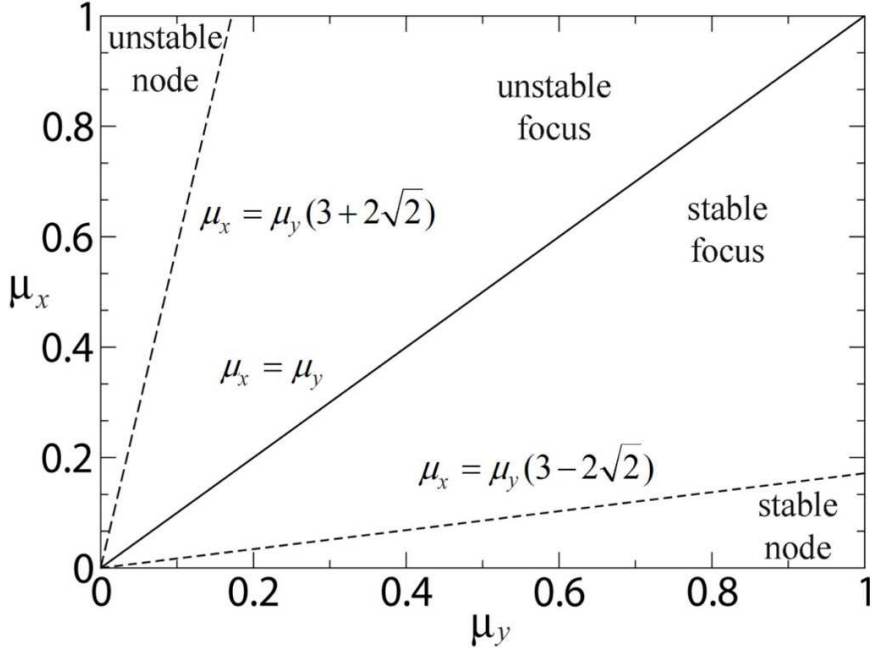}
\caption{\label{eigen} Stability of the uncoupled equilibrium points of the Meinhardt-Gierer equations in the parameter plane $\mu_x$ {\it versus} $\mu_y$.}
\end{figure}

The linearized dynamics around this equilibrium, for the uncoupled case, is obtained from the Jacobian matrix of (\ref{mg1})-(\ref{mg2}), with elements evaluated at the point $(x^*,y^*)$, giving the matrix elements
\bea
\label{ja}
a & = & \mu_x, \qquad
b = -\frac{\mu_x^2}{\rho_x}, \\
\label{jd}
c & = & \frac{2 \rho_x \mu_y}{\mu_x}, \qquad
d = - \mu_y, 
\eea
\no If the equilibrium  $(x^*,y^*)$ is to be stable, the trace and determinant of this Jacobian must be, respectively, negative and positive, yielding the following conditions
\beq
\label{condest}
\mu_x < \mu_y, \qquad \mu_x \mu_y > 0
\eeq
\no Since all coefficients are supposed positive, this implies that the decay rate of the inhibitor must be greater for the activator, which is a reasonable assumption for having a stable equilibrium concentration of both species. The approach to the equilibrium is determined by the nature (real or complex) of the eigenvalues of the Jacobian. The results of this simple analysis are summarized in Fig. \ref{eigen}. By the same token, the equilibrium at the origin can be shown to be always unstable and thus it is not interesting when discussing Turing instability due to diffusion.

Now we turn into the coupled system (\ref{chainx})-(\ref{chainy}), for which there is an spatially homogeneous equilibrium state given by 
\beq
x_k = x^*, \qquad y_k = y^*, \qquad (k=1,2,\cdots N)
\eeq
\no This state defines a two-dimensional invariant subspace embedded into the full $2N$-dimensional phase space of the coupled oscillator chain. The subspace is invariant for any trajectory which originates from an initial condition belonging to this subspace is bound to lie within it for further times. Geometrically speaking, the possible spatially non-homogeneous states one may find represent trajectories outside this invariant subspace.

In this geometrical language, we are interested to investigate the stability of the invariant homogeneous subspace along the $2(N-1)$ remaining transversal directions. We can use the results of Section II for obtaining analytical conditions for the stability of the spatially homogeneous state. Substituting (\ref{ja})-(\ref{jd}) into (\ref{auxvar}) there results that the stability conditions (\ref{cond1})-(\ref{cond3}) read
\begin{align}
\label{condg1}
\frac{\mu_x \mu_y}{D_x D_y} & >  0, \\
\label{condg2}
\mu_x D_y - \mu_y D_x > 2 \sqrt{\mu_x \mu_y D_x D_y}, \qquad  & {\mbox{\rm if}} \qquad 0 \le \mu_x D_y - \mu_y D_x \le 4 \sigma_{max} D_x D_y, \\
\label{condg3}
\mu_x D_y - \mu_y D_x > \frac{\mu_x \mu_y}{2 \sigma_{max}} + 2\sigma_{max} D_x D_y, \qquad & {\mbox{\rm if}} \qquad \mu_x D_y - \mu_y D_x > 4 \sigma_{max} D_x D_y, 
\end{align}
\no Since the uncoupled oscillators are supposed to be locally stable, so that Eq. (\ref{condest}) holds, the condition (\ref{condg1}) is always fulfilled, since the diffusion coefficients are positive-definite. 

\begin{figure}
\includegraphics[width=0.7\columnwidth,clip]{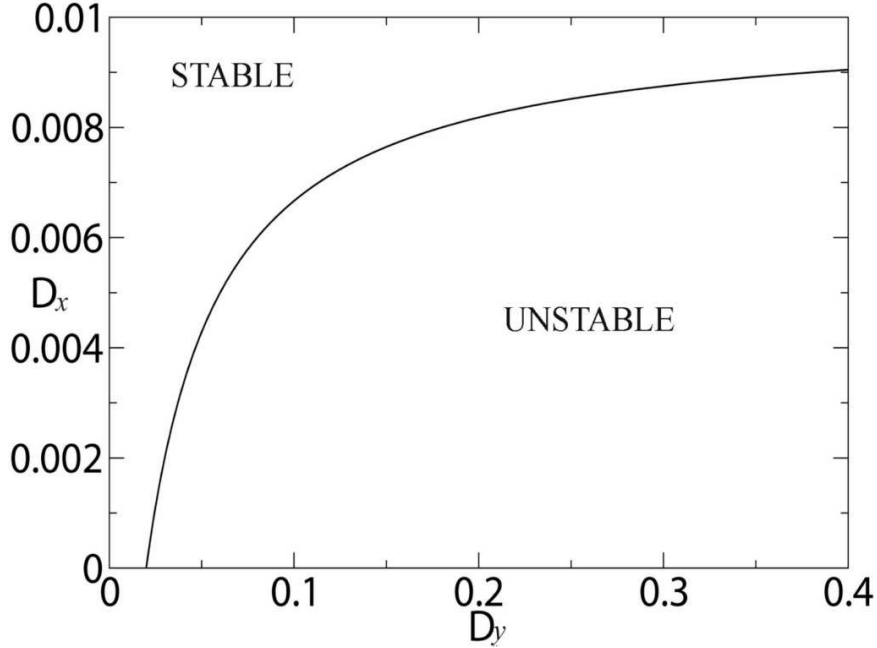}
\caption{\label{pplane} Parameter plane of the activator and inhibitor diffusion coefficients for $\mu_x = 0.01$, $\mu_y = 0.02$, $\rho_x = 0.01$, and $\rho_y = 0.02$.}
\end{figure}

Let us consider the limit cases of this coupling. For nearest-neighbor case (large $\alpha$) we have $\sigma_{max} = 1$. If the diffusion coefficients were equal, there would be no Turing instability, as shown before. For the all-to-all coupling ($\alpha=0$) the conditions (\ref{condg2})-(\ref{condg3}) reduce to the single inequality
\beq
\label{condgl}
\mu_x D_y - \mu_y D_x > \mu_x \mu_y + D_x D_y
\eeq
\no The threshold for Turing instability is depicted in the parameter plane $D_x$ {\it versus} $D_y$ (for fixed values of the remaining parameters) [Fig. \ref{pplane}].

\subsection{Pattern formation}

In the numerical simulations to be presented, we use a predictor-corrector routine (LSODA) based on the Adams method \cite{lsoda}, and the same values for the parameters of the Meinhardt-Gierer model as those chosen in Ref. \cite{koch}: $\mu_x = 0.01$, $\mu_y = 0.02$, $\rho_x = 0.01$, and $\rho_y = 0.02$, for which the equilibrium point is a stable focus with coordinates $(x^*,y^*)=(1.0,1.0)$. We also fix the diffusion coefficient of the inhibitor at $D_y = 0.2$ and use the activator coefficient $D_x$ as the tunable parameter for each value chosen for the effective range $\alpha$. We use, as an initial condition, an almost homogeneous spatial profile with a tiny bump which is intended to be a seed for some unstable mode to grow, after a Turing instability occurs.

\begin{figure}
\includegraphics[width=0.45\columnwidth]{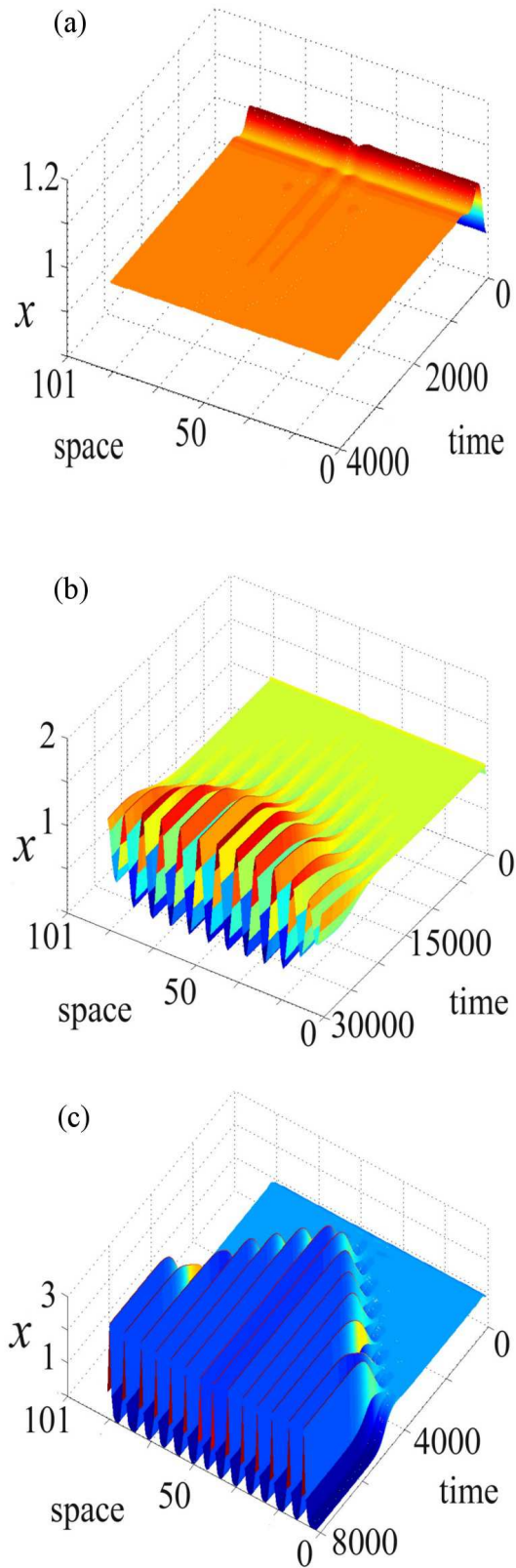}
\caption{\label{localfig} (color online) Space-time diagrams of the activator concentration for the nearest-neighbor coupling case, $D_x = 0.2$, and (a) $D_y = 0.019$; (b) $D_y = 0.016$; (c) $D_y = 0.005$.}
\end{figure}

The local case (large $\alpha$) is illustrated by Fig. \ref{localfig}, where we depict space-time diagrams (activator concentration $x$ {\it versus} discrete space $k$ and time $t$) for different values of the activator diffusion coefficient. For relatively large $D_x$ [Fig. \ref{localfig}(a)] we do not observe pattern formation, since we are ahead from the instability threshold predicted by the linear theory. The oscillators go together to the homogeneous equilibrium state $x^*=1$ instead. The instability threshold occurs for $D_x \approx 0.016$ and we accordingly observe the emergence of a spatially non-homogeneous pattern just after the Turing instability [Fig. \ref{localfig}(b)].

This pattern is roughly a sinusoidal spatial profile with $12$ maxima, corresponding to a wave-length of $\lambda \approx 8.42$ and thus to a wave number
$$
\frac{s}{N} = \frac{1}{\lambda} \approx 0.119, 
$$
\no From the linear stability analysis, the parameter set used to generate Fig. \ref{localfig}(b) corresponds to $\sigma_- = 0.09$ and $\sigma_+ = 0.17$. Using Eq. (\ref{range}), the ranges of unstable modes are $[0.097, 0.135]$ and $[0.865, 0.903]$. The pattern observed has wave number roughly in the middle of the interval of unstable modes, i.e. it is the most unstable mode from the linear approximation. 

As we decrease substantially the activator diffusion coefficient, we observe another spatial pattern [Fig. \ref{localfig}(c)] with $14$ maxima and thus a smaller wavelength than the previous example, namely 
\[
\frac{s}{N}  = \frac{1}{7.21} \approx 0.139,
\]
\no which belongs again to one of the intervals of linearly unstable modes: $[0.076, 0.392]$ and $[0.608, 0.924]$. It is no longer the most unstable mode, however. A larger wave number such as this, is actually a general feature of spatio-temporal systems when we decrease the diffusion coefficient: modes with small wave number are more easily damped when diffusion increases, as a general rule.

\begin{figure}
\includegraphics[width=1.0\columnwidth]{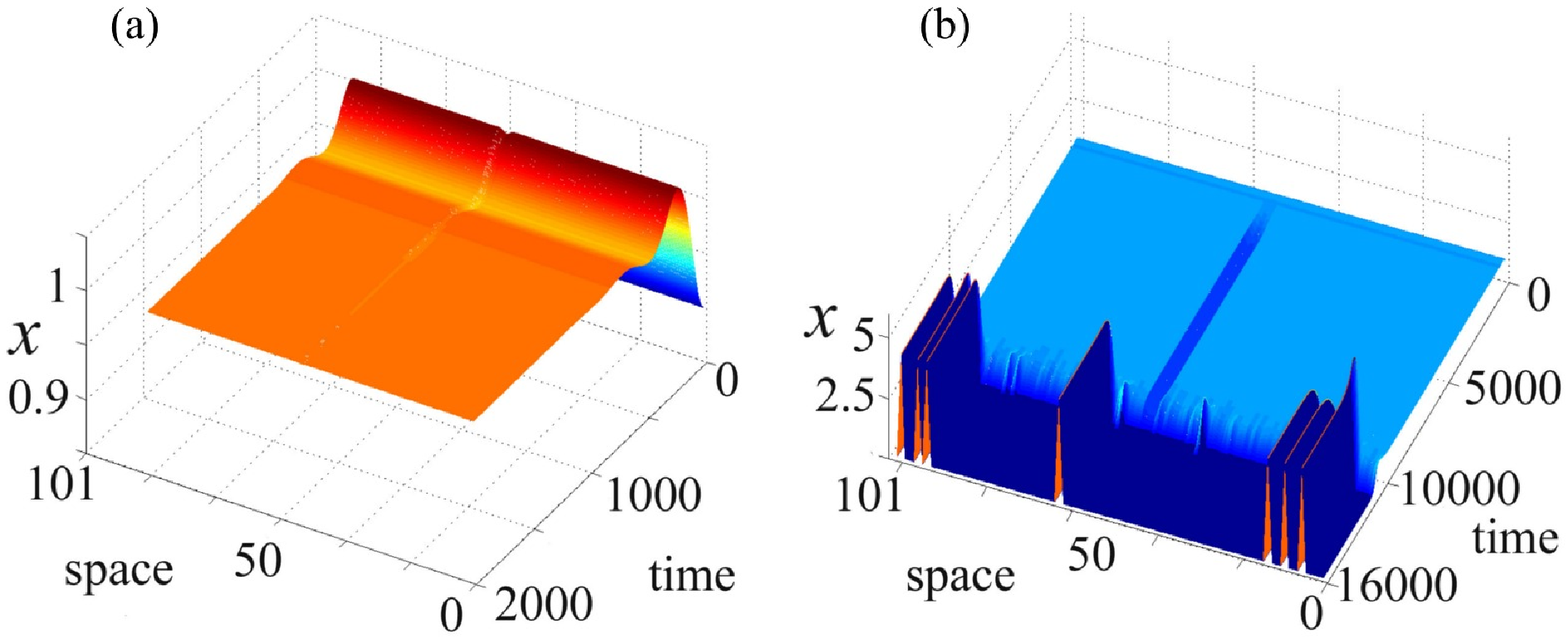}
\caption{\label{globalfig} (color online) Space-time diagrams of the activator concentration for the all-to-all (global) coupling case, $D_x = 0.2$, and (a) $D_y = 0.009$; (b) $D_y = 0.005$.}
\end{figure}

\begin{figure}
\includegraphics[width=1.0\columnwidth]{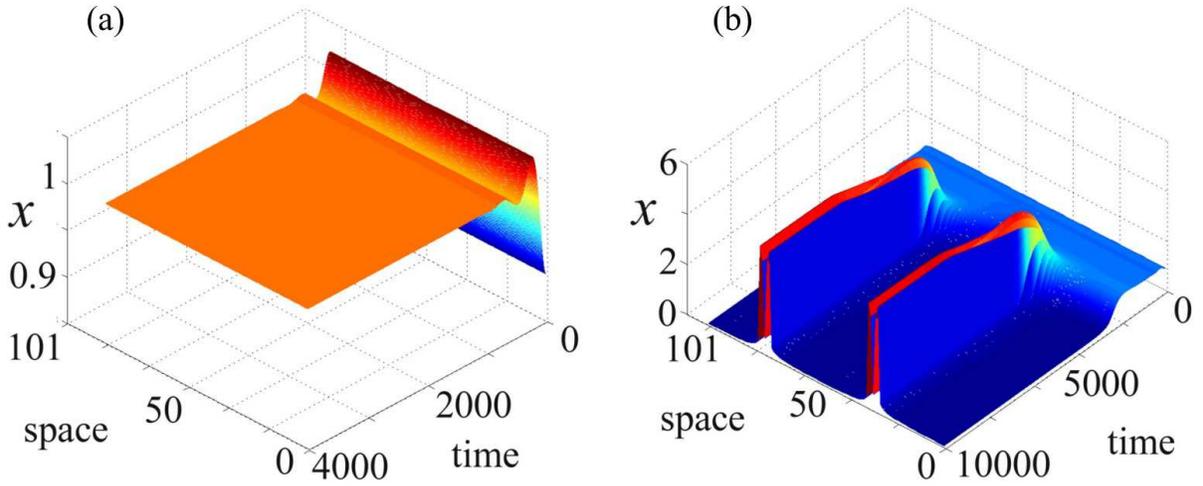}
\caption{\label{alphafig} Space-time diagrams of the activator concentration for the intermediate range coupling of $\alpha = 1.0$, $D_x = 0.2$, and (a) $D_y = 0.020$; (b) $D_y = 0.005$}
\end{figure}

The corresponding space-time plots for the global (all-to-all) coupling case are shown in Fig. \ref{globalfig}(a) and (b) for parameters before and after the occurrence of a Turing instability, respectively. According to the linear theory [see Fig. \ref{pplane}] it occurs for $D_x \approx 0.008$. The pattern formed after the Turing instability is more complex than in the previous example and cannot be assigned to a single unstable mode, but rather to a entire spectrum of unstable modes excited by the Turing instability, and which became stabilized by the saturating effect of the nonlinear terms of the model, a feature obviously not present in the linear stability analysis of the previous Section. The space time plots for the intermediate range coupling characterized by $\alpha = 1.0$ are depicted in Fig. \ref{alphafig}(a) and (b), before and after the threshold of a Turing instability, respectively. 

\section{Conclusions}

Turing instability is a general mechanism of pattern formation, from which a spatially homogeneous pattern becomes unstable and develops a spatial structure after saturation that comes from nonlinear effects. In biological scenarios there may be the need for nonlocal couplings, or interactions among distant cells, not only the nearest neighbors, as in usual studies of spatial diffusion. As far as we know, this is the first analytical study of the occurrence of Turing instability in coupled oscillator chains with nonlocal interactions. We presented a linear stability analysis for the Turing instability when the coupling among sites is non-local, i.e. it takes into account not only the nearest-neighbors but also all the other oscillators in a one-dimensional chain, the coupling strength decreasing with the lattice distance as a power law. 

We shown, using a specific nonlinear model, that the linear stability analysis is useful to characterize the Turing instability when nonlocal couplings are considered. In particular, we derived analytical conditions for the presence of Turing instability for an arbitrary effective coupling range. Our results agree, in the limit of short range, with results obtained for nearest-neighbor couplings, and can be also applied to situations with global (all-to-all) couplings. We shown that local couplings have a comparatively shorter stable area in the parameter plane than global couplings. In other words, as the effective coupling range increases, the more stable are the spatial configurations and, statistically speaking, less probable would be the occurrence of a Turing instability. 

This fact can be qualitatively understood by noting that global couplings spread information among cells more rapidly than locally coupled ones, for which diffusion happens at a slower rate. Then globally coupled cells are less likely to present Turing instability than locally coupled ones. On the other hand, other dynamical collective phenomena due to interactions, like frequency synchronization \cite{sandro}, short-term memory formation \cite{batista}, bursting synchronization \cite{pontes}, are more likely to occur in the globally than locally coupled chains.

The question of pattern formation, however, is more complex, since the patterns are ultimately determined by nonlinear features of the model not predictable by linear stability analysis. Even so, in some cases it is possible to relate single-mode patterns to the modes which become unstable after a Turing instability. 

\section{Acknowledgments}

This work was made possible by partial financial support of CNPq, CAPES, Funda\cao Arauc\'aria, and RNF-CNEN (Brazilian Fusion Network).


\begin{thebibliography}{99}
\bibitem{turing} A. Turing, Phil. Trans. R. Soc. London, {\bf 237}, 37 (1952).
\bibitem{murray} J. D. Murray, {\it Mathematical Biology}, 2nd. Ed. (Springer, Berlin, 1993).
\bibitem{meinhardt4} H. Meinhardt and M. Klingler, J. Theor. Biol. {\bf 126}, 63 (1987).
\bibitem{lengyel} I. Lengyel, G. R\'abai, and I. Epstein. J. Am. Chem. Soc. {\bf 112}, 4606 (1990); {\bf 112}, (1990).
\bibitem{cross} S. Setayeshgar and M. C. Cross, Phys. Rev. E {\bf 58}, 4485 (1998); {\bf 59}, 4258 (1999).
\bibitem{atlee} E. Atlee-Jackson, {\it Perspectives in Nonlinear Dynamics}, Vol. II (Cambridge University Press, Cambridge, 1991.
\bibitem{lopez} C. Lopez, Phys. Rev. E {\bf 74}, 012102 (2006).

\bibitem{pikowsky} M. Rosenblum and A. Pikovsky, Phys. Rev. Lett., {\bf 92}, 114102 (2004) 
\bibitem{strogatz} K. Y. Tsang, R. E. Mirollo, S. H. Strogatz and K. Wiesenfeld, Physica D {\bf 48}, 102 (1991).
\bibitem{kuramoto} Y. Kuramoto, {\it Chemical Oscillations, Waves, and Turbulence} (Dover Publications, 2003)
\bibitem{rogers} J. L. Rogers and L. T. Wille, Phys. Rev. E {\bf 54}, R2193 (1996).
\bibitem{nosso} S. E. de S. Pinto, S. R. Lopes, and R. L. Viana, Physica A {\bf 303}, 339 (2002).
\bibitem{shnerb} N. Shnerb, Phys. Rev. E {\bf 69}, 061917 (2004).
\bibitem{nicola} E. Nicola, M. Bar, and H. Engel, Phys. Rev. E {\bf 73}, 066225 (2006).
\bibitem{meinhardt1} H. Meinhardt, {\it Models of Biological Pattern Formation} (Academic Press, 1992). 
\bibitem{abramowitz} P. B. Kahn, {\it Mathematical Methods for Scientists and Engineers} (Wiley, New York, 1990).
\bibitem{ital} G. Paladin and A. Vulpiani, J. Phys. A {\bf 25}, 4911 (1994); A. Torcini and S. Lepri, Phys. Rev. E {\bf 55}, R3805 (1997).
\bibitem{meinhardt2} A. Gierer and H. Meinhardt, Kybernetik {\bf 12}, 30 (1972). 
\bibitem{meinhardt3} A. Gierer, Progr. Biophys. Molec. Biol. {\bf 37}, 1 (1981).
\bibitem{lsoda} A. C. Hindmarch, {\it ODEPACK: a systematized collection of ODE solvers,} in Scientific Computing, R. S. Stepleman, M. Carver, R. Peskin,W. F. Ames, and R. Vichnevetsky, Eds., vol. 1 of IMACS Transactions on Scientific Computation (North-Holland, Amsterdam, 1983), pp. 55–64.
\bibitem{koch} H. Meinhardt and J. A. Koch, Rev. Mod. Phys. {\bf 66}, 1481 (1994).
\bibitem{sandro} S. E. de S. Pinto and R. L. Viana, Phys. Rev. E. {\bf 61}, 5154 (2000).
\bibitem{batista} J. C. A. de Pontes, A. M. Batista, R. L. Viana, and S. R. Lopes, Physica A {\bf 368}, 387 (2006).
\bibitem{pontes} J. C. A. de Pontes, R. L. Viana, S. R. Lopes, C. A. S. Batista, and A. M. Batista, Physica A {\bf 387}, 4417 (2008).
\end{thebibliography}
\end{document}